\documentclass[lettersize,journal]{IEEEtran}
\usepackage{algorithmic}
\usepackage{textcomp}
\usepackage{mathtools}
\usepackage{dsfont}
\usepackage{amsfonts}
\usepackage{graphicx}
\usepackage{color}
\usepackage{bbm}
\newcommand{\x}{\pmb{x}}

\newcommand{\TD}{T_{\mathrm{D}}}

\newcommand{\m}{\mathrm{m}}
\newcommand{\diffeff}[1]{D_{\mathrm{e},#1}}

\newcommand{\difft}{D_\mathrm{t}}
\newcommand{\y}{\pmb{y}}
\newcommand{\z}{\pmb{z}}
\newcommand{\rthree}{\mathbb{R}^3}
\newcommand{\origin}{\pmb{0}}

\newcommand{\erfc}{\mathrm{erfc}}
\newcommand{\erf}{\mathrm{erf}}
\newcommand{\diff}{\mathrm{d}}
\newcommand{\norm}[1]{\mathrm{\| #1\|}}
\newcommand{\NMij}{\mathsf{P}_{{ij}}}
\newcommand{\Evei}{\mathsf{E}_{{i}}}
\newcommand{\Sij}{\mathsf{S}_{ij}}
\newcommand{\Tij}{\mathsf{T}_{ij}}
\newcommand{\Probi}[1]{\mathrm{P_{i}}\left(#1\right)}
\newcommand{\Prob}[1]{\mathsf{p}\left(#1\right)}
\newcommand{\ProbS}[1]{\overline{\mathsf{p}}\left(#1\right)}
\newcommand{\ProbSd}[1]{\overline{\mathsf{p}}_\mathrm{d}\left(#1\right)}
\newcommand{\Probd}[1]{\mathsf{p}_\mathrm{d}\left(#1\right)}
\newcommand{\Probse}[1]{\mathsf{p}_\mathrm{s}\left(#1\right)}
\newcommand{\expect}[2]{\mathbb{E}_{#1}\left[ #2 \right]}

\newcommand{\indc}{\mathds{1}}

\newcommand{\ie}{\textit{i.e., }}

\newcommand{\bt}[1]{\mathbf{b}{\left(#1\right)}}

\newcommand{\ball}[2]{\mathcal{B}\left(#1, #2\right)}
\newtheorem{theorem}{Theorem}
\newtheorem{lemma}{Lemma}
\newtheorem{corollary}{Corollary}

\begin{document}
\title{On the Target Detection Performance of a Molecular Communication  Network with Multiple Mobile Nanomachines}

	\author{Nithin V. Sabu
	and Abhishek K. Gupta
	\thanks{ N. V. Sabu and A. K. Gupta are with Indian Institute of Technology Kanpur, Kanpur UP 208016, India (Email:{ \{nithinvs,gkrabhi\}@iitk.ac.in}).}}
\maketitle
%%
%% The abstract is a short summary of the work to be presented in the
%% article.
\begin{abstract}
A network of nanomachines (NMs) can be used to build a target detection system for a variety of promising applications. 
They have the potential to detect toxic chemicals, infectious bacteria, and biomarkers of dangerous diseases such as cancer within the human body. Many diseases and health disorders can be detected early and efficiently treated in the future by utilizing these systems. To fully grasp the potential of these systems, mathematical analysis is required. 
This paper describes an analytical framework for modeling and analyzing the performance of target detection systems composed of multiple mobile nanomachines of varying sizes with passive/absorbing boundaries. We consider both direct contact detection, in which NMs must physically contact the target to detect it, and indirect sensing, in which NMs must detect the marker molecules emitted by the target. The detection performance of such systems is calculated for degradable and non-degradable targets, as well as mobile and stationary targets. The derived expressions provide various insights, such as the effect of NM density and target degradation on detection probability.
\end{abstract}

%
% The code below is generated by the tool at http://dl.acm.org/ccs.cfm.
% Please copy and paste the code instead of the example below.
%

%%
%% Keywords. The author(s) should pick words that accurately describe
%% the work being presented. Separate the keywords with commas.
%% Keywords. The author(s) should pick words that accurately describe
%% the work being presented. Separate the keywords with commas.
%\keywords{Mobile nanomachines, target detection and sensing.}
%% A "teaser" image appears between the author and affiliation
%% information and the body of the document, and typically spans the
%% page.

%%
%% This command processes the author and affiliation and title
%% information and builds the first part of the formatted document.

\section{Introduction}
Molecular communication holds the potential for facilitating energy-efficient communication among nanomachines (NMs) by employing molecules as carriers of information \cite{nakano2013a,nakano2012}. Among various propagation mechanisms in molecular communication, diffusion-based propagation has received extensive attention due to its mathematical simplicity, energy efficiency, and minimal communication infrastructure requirements \cite{kuran2010}.  Molecular communication finds application in a wide range of fields, from targeted drug delivery to environmental sensing.

One promising application of molecular communication lies in target detection systems, where the primary objective is to identify harmful or poisonous molecules and infectious micro-organisms \cite{okaie2016a}. Consider a scenario in which we aim to detect a harmful micro-organism within the human body. These target micro-organisms often express specific surface proteins. NMs can be introduced into the same medium and employed in a direct sensing system, making direct contact with the target micro-organism. Another intriguing scenario involves targets emitting distinctive markers. In this case, NMs can detect these markers, indirectly confirming the presence of the target, resembling an indirect sensing approach.
Such systems hold significant potential, particularly in the early-stage detection of cancerous cell biomarkers. It's worth noting that deploying multiple NMs is often necessary in the target environment to achieve adequate coverage, given the limited functionality of a single NM \cite{nakano2013a}. To gain insights into their performance, it is essential to develop mathematical models and conduct comprehensive analyses of these systems.

In recent literature \cite{nakano2017a,okaie2016a,mosayebi2018}, there have been notable investigations into molecular communication-based target detection. For instance, authors in \cite{nakano2017a} introduced a leader-follower-based model in their work, for mobile bio-nanomachines for detecting targets.
In \cite{mosayebi2018}, the authors conducted a study  on a target detection system involving stationary NMs for identifying targets that continuously emit molecules into the surrounding medium.
In our prior research \cite{sabu2020b}, we presented the derivation of the detection probability of a mobile target molecule interacting with multiple stationary NMs. These stationary NMs were equipped with fully-absorbing boundaries and distributed spatially following a Poisson point process (PPP) \cite{andrews2023a,haenggi2012}. 
It is worth noting that  PPP can be used to model the spatial distribution of NMs to include the randomness into their spatial positions as employed in recent molecular communication literature \cite{sabu2019, deng2017a,dissanayake2019a}. The above mentioned works deal with the systems with stationary NMs. Further, \cite{okaie2016a} presented an initial study to explore a target detection and tracking system with stationary and mobile NMs in a two-dimensional context using both zero-dimensional stationary and mobile NMs.
 \cite{cerny2008} has derived the capacity functional of the boolean process formed by a set of moving Brownian particles  forming a PPP. 

However, it's essential to highlight that, as of now, there is a lack of systematic and comprehensive study of molecular detection systems in three-dimensional space that may involve moving NM sensing agents of various sizes to detect moving or stationary targets with a possibility of target degradation.

In this work, we focus on a target detection application aimed at identifying both stationary and mobile targets using a network of multiple mobile NMs of varying sizes acting as sensing agents. Key features of the considered model include equipping the NMs with passive/absorbing boundaries and deploying them randomly within the same environment as the target. Our investigation considers scenarios involving direct contact and indirect sensing for target detection.
In direct sensing, the target is detected when NMs make contact with the target molecule, whereas indirect sensing involves the detection of markers emitted by the target.
We develop an analytical framework to study these systems comprehensively. 
Specifically, our modeling approach involves representing the centers of the NMs as a PPP within the medium, with the NMs undergoing Brownian motion in three-dimensional space. The analysis accounts for both degradable and non-degradable target molecules along with the mobility of the target. 

%The derived equations for target detection systems with multiple NMs are valid for both passive and absorbing receivers since we focus on the probability that any of the NMs detect or sense the target. 

Our contributions in this work can be summarized as follows:
\begin{itemize}
	\item We begin by presenting the detection probability of a target molecule when interacting with a single mobile NM. This analysis covers scenarios with degradable, non-degradable, mobile, and stationary target molecules.
	\item We then proceed to derive the detection probability of a stationary target molecule when multiple mobile NMs of varying sizes are deployed in the medium. Our analysis accounts for both degradable and non-degradable target molecules. The initial deployment of NMs follows a uniform PPP distribution.
	\item We extend our investigation to scenarios involving a mobile target molecule and multiple mobile NMs, considering both degradable and non-degradable targets. Along with the detection probability, we also derive the mean detection time, representing the average duration for the NMs to successfully detect a target molecule, whether stationary or mobile, within a given environment. 
	\item  Finally, we explore an indirect sensing system where NMs are deployed to sense the target's presence by monitoring the concentration of marker molecules emitted by the target.  We derive the sensing probability for such a system at time $t$ and within time $t$.
	\item In comparison to particle-based simulations, which take several hours, the use of the derived analytical expressions reduces the simulation time to a few seconds. We also numerically compare various detection systems presented in this work.

\end{itemize}

\textbf{Notation :} $ \ball{\x}{a} $ represents a ball of radius $ a $ centered at the
location $ \x $. $ \|\x\| $ denotes the Euclidean norm of the vector $ \x $. $ A \oplus B $ represents the Minkowski sum of the two sets $ A $ and $ B $. $ \mid A\mid $ is the volume of $ A $. $ \phi $ denotes null set.
\begin{figure*}[ht]
	\centering
	\includegraphics[width=\linewidth]{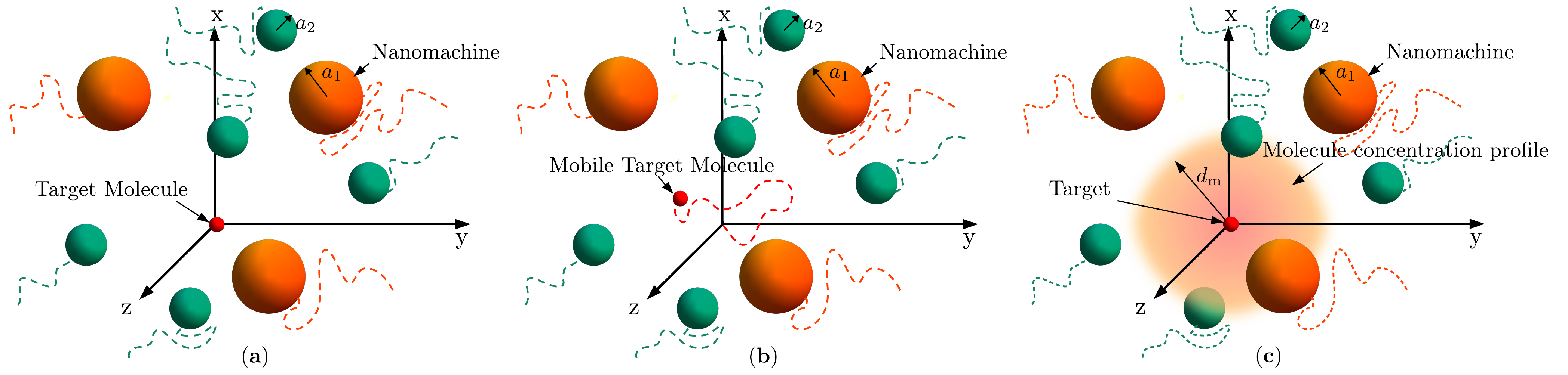}
	\caption{Target detection systems with (a) stationary target and PPP distributed NMs based on direct contact based detection, (a) mobile target and PPP distributed NMs based on direct contact based detection, and (c) PPP distributed NMs based on indirect sensing. }
	\label{fig:sm}
	
\end{figure*}
\section{System Model}
In this section, we present the system model for various configurations of target detection systems operating in a three-dimensional space ($ \mathbb{R}^3 $). The systems are designed to detect stationary or mobile targets using a network of mobile NMs of different sizes acting as sensing agents. We assume that each NM possesses a spherical geometry with passive/absorbing boundaries. The formulated equations for these systems are applicable to NMs of both passive and absorbing boundaries, given that the event of interest occurs when the NM boundary contacts a molecule for the first time. Due to the small size of the NMs and the absence of any other movement mechanism, their motion can be modeled as Brownian. We assume that the target is situated at the origin.

Along with non-degradable targets, we also consider the scenario where the target molecule undergoes degradation over time due to interactions with other molecules in the medium. In this context, the NM can only detect the target molecule if the NM hits the molecule before degradation occurs. We assume that the degradation of the target molecule follows a first-order degradation process \cite{heren2015a}. Therefore, the probability that the target molecule does not degrade within time $ t $ is represented as $ \exp\left(-\mu t\right) $, where $ \mu $ denotes the degradation rate constant. In mathematical terms, this can be expressed as:

\begin{align}
	\mathbb{P}\left[t_{\diff}>t\right]=e^{-\mu t}, \ t\geq 0,\label{degeq}
\end{align}
where $ t_{\diff} $ signifies the degradation time.

We further assume that the target and mobile NMs
do not interact with each other. Hence, the paths of the NMs
remain unaffected by the target.

In this paper, we consider three configurations of target detection systems.

 \textit{C-1. Detection of a Stationary Target by Mobile NMs.}
Here, we consider multiple mobile NMs deployed for target detection. 
 There are $n$ types of NMs with $i$th type having radius $ a_i,\ i=1,2,\cdots,n $, and diffusion coefficient $ D_i,\ i=1,2,\cdots,n $. Initially (\ie at time $t=0$), the NMs of $i$th type are distributed as a homogeneous PPP with density $ \lambda_i$, outside a spherical region of radius $ r $. This configuration is especially relevant when deploying NMs proximate to the target is impractical. It is important to note that $ r\geq \max_i{a_i} $. It is assumed that these NMs do not interact with each other.

The NMs with radius $ a_i $, collectively form a PPP $ \Phi_i $ with a density $ \lambda_i (\x)=\lambda_i\mathbb{I}\left(\x\notin \ball{\origin}{r}\right)$ at time $t=0$. Their centers are denoted by $ \x_{ij},\ j\in\mathbb{N}$'s.  The union of $\Phi_i$'s (\ie  $ \Phi_\mathrm{s}=\cup_{i=1}^{n} \Phi_i$) forms a PPP with a density $ \lambda(\x)= \sum_{i=1}^{n} \lambda_i (\x)$ denoting the location of all NMs at time $t=0$. See Fig. \ref{fig:sm} (a).

 Let $ b_{ij}(t) $ represent the relative Brownian path (path taken by the NM with respect to its starting point) of the NM initially located at $ \x_{ij} $ (denoted as $ \NMij $) during time $ t $. Consequently, the actual path of $ \NMij $ is $\bt{\x_{ij},t}= \x_{ij}+ b_{ij}(t)$. We define $ \Sij(t)= b_{ij}(t) \oplus \ball{0}{a_i} $ as a mark of the NM at $ \x_{ij} \in \Phi_i $. It signifies the region covered by the NM in time duration $ [0,t] $ relative to its center. Hence, the actual region covered by this NM is $\Sij(\x_{ij},t) =\x_{ij} +\Sij(t) $. Detection of the target by $ \NMij $ in time duration $ [0,t] $ occurs when the target intersects with the region $ \Sij(\x_{ij},t) $. 
% In this direct contact based detection, the target molecule is detected only when any of the NM comes in contact with the target molecule initially located at the origin. The target molecule could represent various entities, such as a poisonous molecule or a biomarker originating from a cancer cell. 

\textit{C-2: Detection of a Mobile Target by Multiple Mobile NMs.} 
We then consider a system where the target also undergoes Brownian motion with the initial location at the origin. (See Fig. \ref{fig:sm} (b)). The rest of the model remains the same as C-1.

\textit{C-3: Indirect Sensing of a Stationary Target by Multiple Mobile NM.} 
We also consider an indirect sensing scenario where the target, such as a cancer cell located at the origin, emits molecules (biomarkers) into the surrounding environment. These markers propagate in the medium via Brownian motion with diffusion coefficient $D$. This model is similar to the abovementioned model; however, the NMs do not need to detect the target itself directly. Instead, they can sense the presence of the target by assessing the concentration of molecules emitted by it as in Fig. \ref{fig:sm} (c). If the concentration of molecules at any location surpasses a predefined threshold value $\eta$, then their presence can be sensed by an NM present at the same location. This approach is particularly valuable when target molecules degrade rapidly while leaving a trace. 

\section{Stationary Target Detection via Direct Contact using Mobile Nanomachines}

In this section, we present the detection probabilities of a target molecule when one or more NMs are deployed in the medium. In our setup, we consider a stationary or mobile target molecule initially located at the origin of the Cartesian coordinate system, while single or multiple mobile NMs are spatially deployed in the same medium to detect it. Here, we consider direct contact detection, where the NMs must physically reach the target molecule to detect its presence successfully.

\subsection{Single Mobile Nanomachine Case}\label{Sec:singleNM}

We first consider a scenario where a mobile NM is employed to detect a stationary target molecule. 

Let us consider that a single mobile NM with a radius of $a_1$ is initially located at $\x_1$ at time $t = 0$ with diffusion coefficient $D_1$. 
Recall that with respect to the starting point, $b_1(t)$ represents the Brownian path of NM with respect to its starting point in time $t$ and $ \mathsf{S}_1(t)= b_{1}(t) \oplus \ball{0}{a_1} $ represents the region covered. The actual path is $\bt{\x_{1},t}= \x_{1}+ b_{1}(t)$ and the actual region covered is $\mathsf{S}_1(\x_{1},t) \stackrel{\Delta}{=}\x_{1} +\mathsf{S}_1(t) =\bt{\x_{1},t}\oplus \ball{0}{a_1}$. In direct contact-based detection, the target molecule is detected only when any of the NM comes in contact with the target molecule. Hence, Detection of the target by $\mathsf{P}_1$ in time duration $[0, t]$ occurs when the target intersects with the region $\mathsf{S}_1(\x_{1},t)$.

Hence, the event of detecting a stationary target at the origin via this mobile NM is equivalent to the condition that the origin lies within $ \mathsf{S}_1(\x_{1},t) $ \ie $ \origin\in \mathsf{S}_1(\x_{1},t)= \bt{\x_1,t} \oplus \ball{\origin}{a}$. Further, this event is equivalent to the event $\bt{\x_1,t} \cap \ball{\origin}{a} \neq \emptyset $, signifying the event of a mobile molecule starting from $\x_1$ intersecting a stationary receiver with a radius of $ a_1 $ at the origin. The probability of this event was given in \cite{yilmaz2014a}. 

Hence, the detection probability $ \ProbS{t} $ of a non-degradable target molecule at the origin by an NM \textit{within time $ t$} is given as

\begin{align}
	\ProbS{t}=\frac{a_1}{d_1}\erfc\left(\frac{d_1-a_1}{\sqrt{4D_1 t}}\right),\label{singlenm}
\end{align}
where $ d_1=\norm{\x_1} $  denotes the NM's initial distance from the target.

Now, we consider a scenario where the target molecule undergoes degradation over time due to interactions with other molecules in the medium. Recall that $ t_{\diff} $ signifies the degradation time. We define $ \ProbSd{t} $ as the detection probability of a degradable target molecule at the origin by the NM within time $ t $, which is given by

\begin{align}
	\ProbSd{t}&=\int_{0}^{t} \frac{\diff \ProbS{\tau}}{\diff\tau} \times \exp (-\mu \tau) \diff \tau.
\end{align}
Here $\ProbS{\tau}$ denotes the probability that NM intersects the target first time in duration $[\tau,\tau+\diff \tau]$, and $\exp(-\mu t)$ is the probability that it is not degraded yet. Substituting the value from \eqref{degeq}, we get
\begin{align}
	\ProbSd{t}&{=}\frac{a_1}{2d_1}\left[\exp \left(-\sqrt{\frac{\mu}{D_1}}(d_1-a_1)\right) \erfc\left(\frac{d_1-a_1}{\sqrt{4 D_1 t}}{-}\sqrt{\mu t}\right)\right. \nonumber\\
	&\quad\left.+\exp \left(\sqrt{\frac{\mu}{D_1}}(d_1-a_1)\right) \erfc\left(\frac{d_1-a_1}{\sqrt{4 D_1 t}}+\sqrt{\mu t}\right)\right]
	.\label{singledegnm}
\end{align}
As discussed above, this is the same as the probability that a degradable mobile molecule comes in contact with the surface of a stationary NM within time $t$ \cite{heren2015a}.
Note that when $\mu=0$, this detection probability reduces to  \eqref{singlenm}. 

We now extend the analysis for systems with multiple NMs.

\subsection{Detection of a Stationary Target Molecule by a Network of Mobile Nanomachines}

As the NMs move in the space, their locations  $ b(\x_{ij},t)$ form a point process at any given instant. We denote $ \Psi_i(t) $ as the point process containing the locations of the $i$-type NMs at time $ t $, that is, $ \Psi_i(t)=\cup_{\x_{ij}\in\Phi_i} b(\x_{ij},t) $. According to the displacement theorem \cite{andrews2023a}, if $ \Psi_i(t) $ is a PPP, then $ \Psi_i(t+dt) $ is also a PPP. Since $\Psi_i(0)=\Phi_i$ is a PPP, $ \Psi_i(t) $ remains a PPP for all $ t $. The density of $\Psi_i(t)$ is given as \cite{andrews2023a}
\begin{align}
\lambda_{\Psi_i(t)}(\y)=\lambda_i \int_{\mathbb{R}^3/\ball{\origin}{r})} p(\x-\y,t) \diff \x\label{Psidens}
\end{align}
where $p(\z,t)= \frac{1}{(4\pi D_it)^{3/2}} e^{-\frac{||\z||^2}{4D_it}}$.

 Note that, when $\y=0$,
\begin{align}
\lambda_{\Psi_i(t)}(0)&=4\pi \lambda_i \int_{r}^{\infty} \frac{1}{(4\pi D_it)^{3/2}} e^{-\frac{r^2}{4D_it}} \diff r\nonumber\\
&=\lambda_i\erfc\left(\frac{r}{\sqrt{4D_it}}\right)+\frac{\lambda_i 2r}{\sqrt{4\pi D_it}}\exp\left(-\frac{r^2}{4D_it}\right).
\end{align}

It's worth noting that $ \Phi_i=\Psi_i(0) $.

%Furthermore, we define $ \Xi_i(t) $ as the process that encompasses all the locations traveled by the center of the NMs in $ \Phi_i $ during the time interval $ [0,t] $, which can be expressed as $ \Xi_i(t)=\cup_{\x_{ij}\in\Phi_i} \bt{\x_{ij},t} $. Additionally, we denote $ \Xi^{(a_i)}i(t) $ as the process comprising all the locations covered by the NMs in $ \Phi_i $ within the time duration $ [0,t]$, which can be defined as $ \Xi^{(a_i)}i(t)=\cup{\x_{ij}\in\Phi_i} \bt{\x_{ij},t} \oplus \ball{0}{a_i}$. Importantly, we can demonstrate that both $ \Xi(t) $ and $ \Xi^{(a_i)}_i $ are boolean processes.
\subsubsection{Non-Degradable Target Molecule} \label{StatNoDeg}
In this scenario, we consider the target to be non-degradable. 
We denote $\Gamma_i$ as the set of NMs $\x_{ij}$'s for which the corresponding region $\Sij(\x_{ij},t)$ intersects the target at the origin, i.e., $\Gamma_i=\{\x_{ij}: \x_{ij} +\Sij(t)\cap\origin\neq \phi\}$.

Since the paths of NMs are independent, $\Gamma_i$ is obtained by independent thinning of the PPP $\Phi_i$, and hence, the derived point process $\Gamma_i$ also follows a PPP distribution. Hence, we have the following result.
\begin{lemma}\label{PPPlemma}
 The number of NMs detecting the target follows a Poisson distribution.
\end{lemma}
Let $N_{\Gamma_i}$ be the number of NMs in $\Phi_i$ passing through the target, i.e., $N_{\Gamma_i}=\mid\Gamma_i\mid=\sum_{\x_{ij}}\indc\left(\x_{ij} +\Sij(t)\cap\origin\neq \phi\right) $.
Now, the mean number of NMs in $ \Phi_i $ that detect the target is given in the following lemma.
\begin{lemma}\label{lmean}
	The mean number of NMs in $ \Phi_i $ that can detect the target molecule at the origin within time $ t $ is given by
	\begin{align}
		\kappa_i(D_i,t)
		&{=}\expect{}{N_{\Gamma_i}}{=}2\pi a_i\lambda_i\left(a_i^2-r^2+2D_it\right)\erfc\left(\frac{r-a_i}{\sqrt{4D_it}}\right)\nonumber\\
		&\quad +4\lambda_i\sqrt{\pi D_it}a_i(r+a_i)\exp\left(-\frac{(r-a_i)^2}{4D_it}\right)\label{nodegmean}
	\end{align}
	Proof:
	See Appendix \ref{amean}.
\end{lemma} 
From Lemma \ref{lmean}, we can observe that the average number of NMs detecting the target molecule increases with time, density, radius, and the diffusion coefficient of the NMs.
The probability that any of the NMs detects the target molecule at the origin within time $ t $ is given in the following theorem.

\begin{theorem}\label{th1}
	The probability that any of the NMs detects the target molecule within time $ t $ is given by
	\begin{align}
		\Prob{t}=&1-\exp\left(-\sum_{i=1}^{n} \kappa_i(D_i,t)\right).\label{ndprob}
	\end{align}
	Proof:
	See Appendix \ref{ath1}
\end{theorem}
%Note that, this theorem is valid for both absorbing and passive NMs because we are interested in detecting the target molecule for the first time when the NM's boundary touches the target molecule.
\begin{corollary}
    When $ r=a_i=a $ ($ D_i=D $ and $ \lambda_i=\lambda $ correspondingly), \ie when all the NMs have the same radius, and they are deployed everywhere in $ \rthree $ (excluding the region of overlap with the target molecule at the origin, \ie $ \ball{0}{a} $), the probability that any of the NMs detects the target molecule within time $ t $ is
	\begin{align}
		\Prob{t}=1-\exp\left(- 4na\pi Dt\lambda-8na^2\lambda\sqrt{\pi D_it}\right).
	\end{align}
\end{corollary}

\subsubsection{Degradable Target Molecule}\label{StatDeg}
Now, we consider the scenario where the target molecule is degradable. Detection of the degradable target within time $ t $ by $ \NMij $ occurs if it is within the region $ \x_{ij} +\Sij(t)$ when $ t_{\diff}>t $. If $ t_{\diff}<t $, then detection by  $ \NMij $ within time $ t $ happens when the target falls within the region $ \x_{ij} +\Sij(t_\diff)$. The corresponding volumes covered by $ \NMij $ are $|\x_{ij}+\Sij(t)|$ when $ t_{\diff}>t $ and $|\x_{ij}+\Sij(t_\diff)|$ when $ t_{\diff}<t $, respectively. In other words, the total volume covered by $ \NMij $ is $|\x_{ij}+\Sij(\min\{t,t_\diff\})|$. Notably, both $ t_\diff $ and $ \x_{ij} $ are stochastic variables, along with the Brownian path. Let $ \rho(D_i,t| t_\diff ) $ denote the total number of NMs detecting the target before its degradation, conditioned on $ t_\diff $ \ie
\begin{align}
\rho(D_i,t\mid t_\diff )&=\lambda_i\expect{\Sij(t)\mid t_\diff}{|\x_{ij}+\Sij(\min\{t,t_\diff\})|} \nonumber\\
&=\left\{\begin{matrix}
 \kappa_i(t_\diff),& \text{ if } t_\diff\leq t \\ 
  \kappa_i(D_i,t),& \text{ if } t_\diff> t 
\end{matrix}\right. \label{rhoi}
\end{align}
\begin{lemma}\label{ldmean}
	The mean number of $i$-type  NMs in $ \Phi_i $ detecting the target molecule at the origin within time $ t $ before its degradation  is given by
	\begin{align}
		&\zeta_i(D_i,t)=\expect{t_\diff}{\rho(D_i,t\mid t_\diff )}=\nonumber\\
		&\quad2\pi\lambda_ia_i\left[\frac{D_i}{\mu}{+}r\sqrt{\frac{D_i}{\mu}}\right]e^{-(r-a_i)\sqrt{\frac{\mu}{D_i}}}\erfc\left(\frac{r-a_i}{\sqrt{4D_it}}{-}\sqrt{\mu t}\right)\nonumber\\
		&\quad+2\pi\lambda_ia_i\left[\frac{D_i}{\mu}{-}r\sqrt{\frac{D_i}{\mu}}\right]e^{(r-a_i)\sqrt{\frac{\mu}{D_i}}} \erfc\left(\frac{r-a_i}{\sqrt{4D_it}}{+}\sqrt{\mu t}\right) \nonumber\\
		&\quad-4\pi a_i\lambda_i\frac{D_i}{\mu}e^{-\mu t}\erfc\left(\frac{r-a_i}{\sqrt{4D_it}}\right).\label{degmean}
	\end{align}
	Proof:
	See Appendix \ref{admean}.
\end{lemma} 
Similar to the non-degradable case, the number of NMs hitting the target is Poisson distributed. Hence, the probability that any of the NMs detect the target at the origin within time $ t $ is given in the following theorem.
\begin{theorem}\label{thdeg}
    The probability that any of the NMs detect the target molecule before its degradation within time $ t $ is given by
	\begin{align}
		\Probd{t}&=1-\expect{t_{\diff}}{\exp\left(-\sum_{i=1}^{n}\rho(D_i,t\mid t_\diff )\right)},\label{th4prob}
	\end{align}
	where $\rho(D_i,t\mid t_\diff )$ is given in \eqref{rhoi}.
	
	Proof: See Appendix \ref{athdeg}.
\end{theorem}
Since further simplification of  \eqref{th4prob} is difficult, we present an approximation using cumulant expansion \cite{hald2000a} in the following result.
\begin{corollary}\label{coronew}
	The probability that any of the NMs detect the target molecule before its degradation within time $ t $ is approximately given as
	\begin{align}
		\Probd{t}\approx 1-\exp\left(-\sum_{i=1}^{n}\zeta_i(D_i,t)\right),\label{approxeq}
	\end{align}
	where $\zeta_i(D_i,t)$ is given in \eqref{degmean}.
	
	Proof:
	See Appendix \ref{athdegcoro}.
\end{corollary}

\begin{corollary}
    The probability that NMs eventually detect the target molecule before its degradation within time $ t $ is given by
	\begin{align}
		\Probd{\infty}\approx& 1-\exp\left(-4\pi\sum_{i=1}^{n}\lambda_ia_i\left[\frac{D_i}{\mu}+r\sqrt{\frac{D_i}{\mu}}\right]\right.\nonumber\\
		&\times \left.e^{-(r-a_i)\sqrt{\frac{\mu}{D_i}}}\right).\label{eqcor3}
	\end{align}
\end{corollary}
It is interesting to note that \eqref{eqcor3} is a function of  $\frac{D_i}{\mu}$, which implies that the effect of an increase in degradation can be taken care of by increasing $D_i$.
\begin{corollary}
    In the case where $ r=a_i=a $ (with $ D_i=D $ and $ \lambda_i=\lambda $), signifying that all NMs share the same radius and are distributed uniformly throughout $ \rthree $ (except for the region overlapping with the target molecule at the origin, i.e., $ \ball{0}{a} $), the probability that any of the NMs detects the target molecule within time $ t $ and before its degradation is given by
	\begin{align}
		\Probd{t}{\approx}& 1{-}\exp\left(-4n\pi a{\sqrt{\frac{D}{\mu}}}\lambda\left(\sqrt{\frac{D}{\mu}}\left(1{-}e^{-\mu t}\right)\right.\right.\nonumber\\
		&\left.+a\erf\left(-\sqrt{\mu t}\right)\bigg)\right).\label{eqra}
	\end{align}
\end{corollary}

\subsection{Mean Detection Time}
Mean detection time (represented by $\sigma_{\TD}$) denotes the time taken by the NMs on average to detect the target molecule.  It serves as a fundamental parameter to evaluate the efficiency and reliability of the target detection system employing a network of mobile NMs of varying sizes. 

Let $\TD$ be the detection time of the target by the NMs. Note that, the cumulative density function of $\TD$, represented by $F_{\TD}(t)$ ($F_{\TD}(t)=\mathbb{P}[\TD\leq t]$), is same as the target detection probability. Therefore, the mean detection time of the target detection system is given in the following corollary.

\begin{corollary}\label{mdt}
The mean detection time of a target detection system with mobile NMs and stationary or mobile target with or without degradation is given by
\begin{align}
\sigma_{\TD}&=\int_0^{\infty}\left(1-F_{\TD}(t)\right)\diff t\nonumber\\
&=\int_0^{\infty}\left(1-p(t)\right)\diff t,
\end{align}
where $p(t)$ is $\Prob{t}$ and $\Probd{t}$ for non degradable and degradable target, respectively.
\end{corollary}

Reducing the mean detection time is often a key objective, as it directly impacts the system's responsiveness and ability to detect targets swiftly, which can be crucial in applications ranging from healthcare to environmental monitoring.

%\begin{corollary}
%	When the NMs are of the same size $ a $, density $ \lambda $ and diffusion coefficient $ D $, the probability that any of the NMs detect the target is
%	\begin{align}
	%		\Prob{t}&=1-\exp\left(-4n\pi t\lambda aD+8na^2\sqrt{\pi Dt}\lambda \right).\label{c1}
	%	\end{align}
%\end{corollary}
%\begin{remark}
%When $ n=1 $, \eqref{c1} gives the probability that any of the NMs in $ \Phi_1 $  detect the target molecule within time $ t $ \cite{manybody,diffcontrl}. \ie
%\begin{align}
%	\Probi{t}=1-\exp\left(-4\pi t\lambda_1 a_1D_1+8a_1^2\sqrt{\pi D_1t}\lambda_1\right).\label{probi}
%\end{align}
%\end{remark}
\section{Detection of a Mobile Target Molecule by a Network of Mobile Nanomachines}
Now, we focus on a system in which both the target and NMs are undergoing Brownian motion.  
 The diffusion coefficient of the target is denoted as $\difft$. Let us consider a single NM case first of type $1$. 
Using the change of reference, we can equate the motion with a system having a stationary target molecule and a mobile NM with an effective diffusion coefficient $\diffeff{1}=\difft+ D_{1}$ \cite{huang2018,nakano2019}. Hence, the
 event of detecting a mobile target with diffusion coefficient $\difft$ and a mobile NM with diffusion
coefficient $D_{1}$ is equivalent to the detection of a stationary target molecule by an NM with an effective diffusion coefficient $\diffeff{1}$. So, in this case, the detection probabilities for the target without and with degradation are given by

\begin{align}
	\ProbS{t}=\frac{a_1}{d_1}\erfc\left(\frac{d_1-a_1}{\sqrt{4\diffeff{1} t}}\right),\label{singlenmDe}
\end{align}
and
\begin{align}
	&\ProbSd{t}{=}\nonumber\\
	&\quad\frac{a_1}{2d_1}\left[\exp \left(-\sqrt{\frac{\mu}{\diffeff{1}}}(d_1-a_1)\right) \erfc\left(\frac{d_1-a_1}{\sqrt{4 \diffeff{1} t}}{-}\sqrt{\mu t}\right)\right. \nonumber\\
	&\qquad\left.+\exp \left(\sqrt{\frac{\mu}{\diffeff{1}}}(d_1-a_1)\right) \erfc\left(\frac{d_1-a_1}{\sqrt{4 \diffeff{1} t}}+\sqrt{\mu t}\right)\right],\label{singledegnmDe}
\end{align}
respectively.

Let us now extend it to the case with multiple NMs. Let $ c_{ij}(t) $ represent the effective path of $j$th $i$-type NM $ \NMij $ within a time interval $ t $ when the NMs move with an effective diffusion coefficient $\diffeff{i}$. Incorporating the effective diffusion coefficient, we can express the modified actual path of $ \NMij $ as $ \bt{\x_{ij},t} = \x_{ij} + c_{ij}(t) $. Furthermore, let $ \Tij(t)= c_{ij}(t) \oplus \ball{0}{a_i} $ denote a mark of the NM at  $ \x_{ij} \in \Phi_i $ that signifies the region covered by it during the time duration $ [0,t] $ relative to its center. The detection of the target occurs when the target at the origin is situated within the region $ \x_{ij} +\Tij(t) $ while $ \NMij $ operates throughout the time interval $ [0,t]$.

The derivation of the mean number of NMs detecting the mobile target is identical to those in sections \ref{StatNoDeg}, and \ref{StatDeg} and the result is provided in the following lemma:
\begin{lemma}
The mean number of NMs in $ \Phi_i $ that can detect a non-degradable and degradable target molecule at the origin within time $ t $ is given by $\kappa_i(\diffeff{i},t)$ and $\zeta(\diffeff{i},t)$, respectively. The total number of NMs that detect the target molecule are hence given by $\sum_i\kappa_i(\diffeff{i},t)$ and $\sum_i\zeta(\diffeff{i},t)$ for non-degradable and degradable target case.
\end{lemma}

Since the target molecule's path is common to the effective paths of all NMs, the paths of all NMs are no longer independent of each other and, hence, no longer jointly Brownian. Consequently, the collection of $\x_{ij}$'s in $\Phi_i$ that intersect with the mobile target is not PPP as in the case of a stationary target due to dependent thinning. We can still approximate it using a PPP to get the approximate probabilities that any of the NMs detect the mobile target as given in the following theorem:

\begin{theorem}
	The approximate probabilities that any of the NMs detect the mobile target molecule within time $ t $ are given as
	\begin{align}
		\Prob{t}&\approx 1-\exp\left(-\sum_{i=1}^{n}\kappa_i(\diffeff{i},t) \right),\\
		\text{and}\nonumber\\
		\Probd{t}&\approx 1-\exp\left(-\sum_{i=1}^{n}\zeta(\diffeff{i},t)\right),
	\end{align}
	for non-degradable and degradable target cases, respectively.
	
	Proof:
	See Appendix \ref{amob}.
\end{theorem}

\section{Indirect Sensing based Detection of a Target with Markers}
We now consider a system where the target continuously emits marker molecules (with diffusion coefficient $D_{\m}$) denoting its presence. In such cases, NMs can sense the concentration of markers to detect the presence of the target indirectly. We assume that NMs do not alter the concentration profile. This case occurs when NMs are either passive or absorbing with very low density and small size.

Recalling that if the target at the origin releases $ N $ molecules impulsively at time $ \tau $, the concentration of molecules at a distance $ d $ at time $ t $ ($>\tau$) is given by \cite{jamali2019}
\begin{align}
	c(d,t)=\frac{N}{\left(4\pi D_{\m} (t-\tau)\right)^{3/2}}\exp\left(-\frac{d^2}{4D_{\m}(t-\tau)}\right).
\end{align}
Considering continuous molecule emission at a rate of $ M(t) $ molecules per second for all time $ t $, the resultant concentration is given by \cite{bossert1963a}
\begin{align}
	f(d,t)=\int_{0}^{t}\frac{M(\tau)}{\left(4\pi D_{\m} (t-\tau)\right)^{3/2}}\exp\left(-\frac{d^2}{4D_{\m}(t-\tau)}\right)\diff \tau.\label{fdt}
\end{align}
Now, if we further assume that emission rate is constant \ie $ M(t)=M $, \eqref{fdt} simplifies to
\begin{align}
	f(d,t)&=\frac{M}{4\pi D_{\m}d}\frac{2}{\sqrt{\pi}}\int_{\frac{d}{\sqrt{4D_{\m}t}}}^{\infty}\exp\left(-x^2\right)\diff x\nonumber\\
	&=\frac{M}{4\pi D_{\m}d}\erfc\left(\frac{d}{\sqrt{4D_{\m}t}}\right).
\end{align}
Supposing that the emission has been occurring for a long time, the concentration becomes
\begin{align}
	f(d,\infty)=\frac{M}{4\pi D_{\m}d},
\end{align}
As a result, the maximum distance $ d_{\m} $ from the target source location up to which the concentration is above a threshold $ \eta $ is given by
\begin{align}
	d_{\m}=\frac{M}{4\pi D_{\m}\eta}.\label{dmax}
\end{align}
Hence, to detect the target's presence, the NMs need to move within $ \ball{\origin}{d_{\m}} $.
It is important to note that we assume that NMs commence detection at time $ t=0 $, and $ d_{\m} $ reaches a steady state before detection is initiated. We now consider two sensing mechanisms in the following two subsections.
\subsubsection{Sensing probability at any time $ t $}
We consider NMs, which measure the concentration at a given time. Hence, sensing probability is defined as the probability that markers' concentration at one of the NMs at a given time $t$ is more than $\eta$. From the discussion above, the sensing probability is equal to the probability that at least one of NM is inside $d_\m$ at time $t$. Recall that at any time $t$, NMs form a PPP with density given in \eqref{Psidens}. Hence, using the void probability of PPP, we can get the sensing probability as given in the following theorem.

\begin{theorem}\label{tsense1}
	The probability of sensing the presence of the target located at the origin by any of the NMs at any time instant is given by:
	\begin{align}
		\Prob{t}=1-\exp\left(- \sum_{i=1}^{n}\int_{\ball{\origin}{a_i+d_{\m}}} \lambda_i(t,\y) \diff \y\right),
	\end{align}
	 where $\lambda_i(t,\y)$ is the same as $\lambda_{\Psi_i(t)}(\y)$ in \eqref{Psidens}.
	 
	Proof: See Appendix \ref{asense1}.
\end{theorem}
%The above derived sensing probability is applicable to NMs with passive boundaries. Note that the passive NMs does not alter the concentration profile of molecules emitted by the target.
\begin{corollary}
	The  probability of detecting the target itself by  any of the NMs at any time instant is given by
	\begin{align}
		\mathrm{p}=1-\exp\left(- \sum_{i=1}^{n}\int_{\ball{\origin}{a_i}} \lambda_i(t,\y) \diff \y\right).
	\end{align}
\end{corollary}
The above corollary gives the detection probability at time $t$ corresponding to a direct contact-based detection system.
\subsubsection{Sensing probability within time $ t $}
The target is sensed by $ \NMij $ within time $ t $, when the volume covered by it (\ie $ \Sij(t) $) intersects with the region with the marker's concentration at least $ \eta $ (which is nothing but a ball of radius $d_\m$ centered around the target, \ie $ \ball{\origin}{d_{\m}} $).

The probability of sensing the target by any of the mobile NMs within time $ t $ is given by the following theorem.
\begin{theorem}\label{th3}
    The probability of sensing the target at the origin within time $ t $  is given by
	\begin{align}
		\Probse{t}&=1-\exp\left(-2\pi\sum_{i=1}^{n} (a_i+d_{\m})\lambda_i\left((a_i+d_{\m})^2-r^2\right.\right.\nonumber\\&\left.\left.+2D_it\right) \times \erfc\left(\frac{r-(a_i+d_{\m})}{\sqrt{4D_it}}\right)-4\sum_{i=1}^{n}\lambda_i\sqrt{\pi D_it}\right.\nonumber\\
		&\left.(a_i+d_{\m})(r+a_i+d_{\m})\exp\left(-\frac{(r-a_i-d_{\m})^2}{4D_it}\right)\right).\label{sensethm}
	\end{align}
	Proof: The target is detected when $ \ball{\origin}{d_\m}\cap\left(\x_{ij} +\Sij(t)\right)\neq \phi $. This is the same as $ \origin\cap\left(\x_{ij} + b_{ij}(t) \oplus \ball{0}{a_i+d_\m}\right)\neq \phi $. Rest of the derivation is similar to that of Appendix \ref{amean} and \ref{ath1}.
\end{theorem}
Comparing \eqref{ndprob} and \eqref{sensethm}, we can verify that the target sensing probability using NMs of radius $ a_i $ is the same as the target detection probability using NMs with radius $ a_i+d_\m $.

\section{Numerical Results}
In this section, we present numerical results to derive insights/understanding from analytical results and validate the accuracy of the derived results. The simulations are conducted using a time step size of $ \Delta t = 10^{-4} $ seconds.
In our simulation setup, NMs of two different sizes are initially deployed as PPP outside a sphere with a radius of $r= 30 \mu m $. This configuration mimics the deployment scenario described in the system model, where NMs are located outside $r$ to infinite space. However, in simulation, we consider NMs inside the ball of $ 150 \mu m $ to ensure a minimal finite simulation window effect. An NM density of $\lambda_i=1\times 10^{-6} \text{NMs}/\mu m^3$ and $\lambda_i=1\times 10^{-5} \text{NMs}/\mu m^3$ corresponds to $14$ and $140$ NMs, respectively in the simulation window.

\begin{figure}
	\centering
	\includegraphics[width=\linewidth]{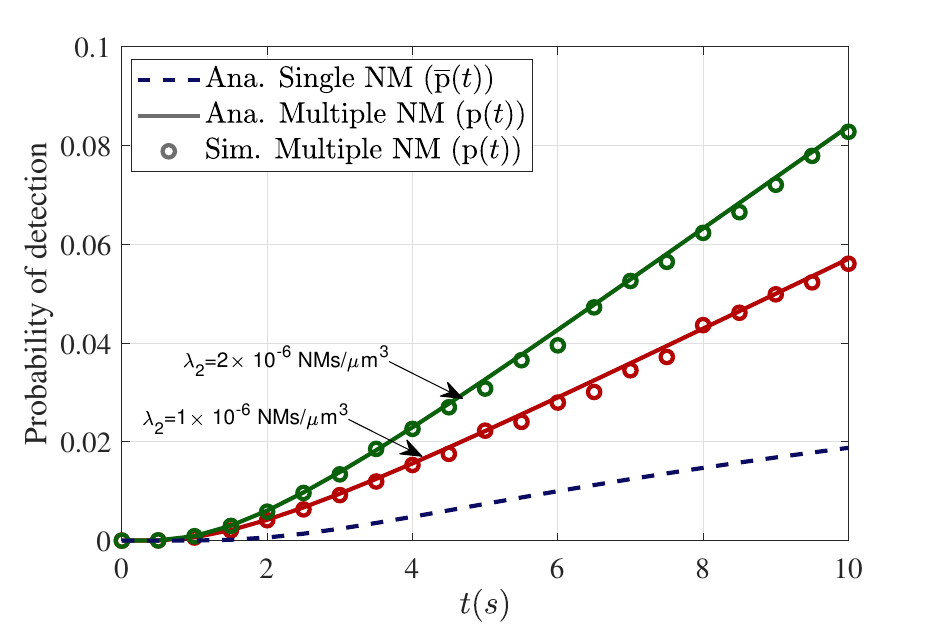}
	\caption{Probability of detecting the non-degradable target versus time for different values of NM density. The value for a single NM case is also shown. Parameters for  single NMs case:  $  a_1=4\mu m,\ D_1=100\mu m^2/s ,\ d=50\mu m$. Parameters for multiple NM case:  $  a_1=3\mu m,\ a_2=4\mu m,\ D_1=100\mu m^2/s,\ D_2=75\mu m^2/s,\ r=30\mu m,\ \text{ and \ } \lambda_1=1\times 10^{-6} \text{NMs}/\mu m^3 $.}
	\label{fig:cdf1}
\end{figure}

 \textbf{ \textit{Impact of Single vs. Multiple NMs on Detection Probability:}} 
 The variation of probability of detection of a non-degradable target within time $t$ with respect to time for various values of NM density is shown in Fig. \ref{fig:cdf1}.  We can see that as the density of the NMs rises, so does the likelihood of detection. Based on \eqref{singlenm}, the dashed line depicts the detection probability for a system with only one mobile NM. We can observe that the use of multiple NMs significantly raises the detection probability when compared to a system with a single NM.

\begin{figure}
	\centering
	\includegraphics[width=\linewidth]{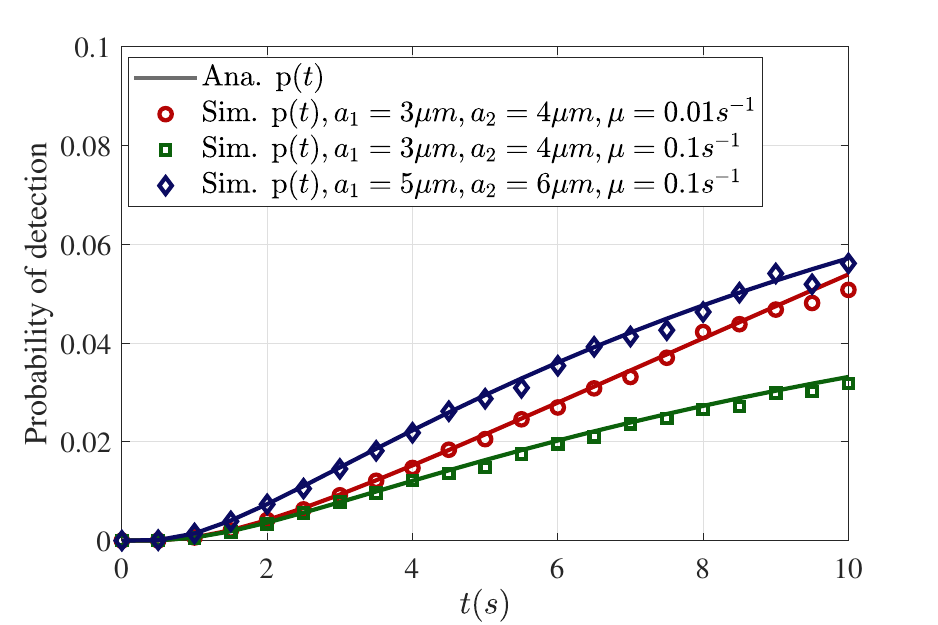}
	\caption{Probability of detecting a stationary target within time $t$ versus time for a degradable target for various values of NM radius and degradation rate. Parameters :  $  D_1=100\mu m^2/s,\ D_2=75\mu m^2/s,\ \text{ and \ } \ \lambda_1=\lambda_2=1\times 10^{-6} \text{NMs}/\mu m^3,\ r=30\mu m$.}
	\label{fig:deg}
\end{figure}
  \textbf{ \textit{Impact of Target Degradation and Size on Detection Probability:}} 
Fig. \ref{fig:deg} shows the variation in the detection probability over time for a degradable target molecule for various values of NM radius and degradation rate. Note that the target molecule breaks down faster as the degradation rate rises. As a result, there is a greater possibility that the target will deteriorate before the NMs get there. The probability of detection declines as a result. Such a negative effect of degradation rate can be countered by deploying NMs with higher radius. Due to the increase in NM's surface area, an increase in NM radius increases the detection probability.

\begin{figure}
	\centering
	\includegraphics[width=\linewidth]{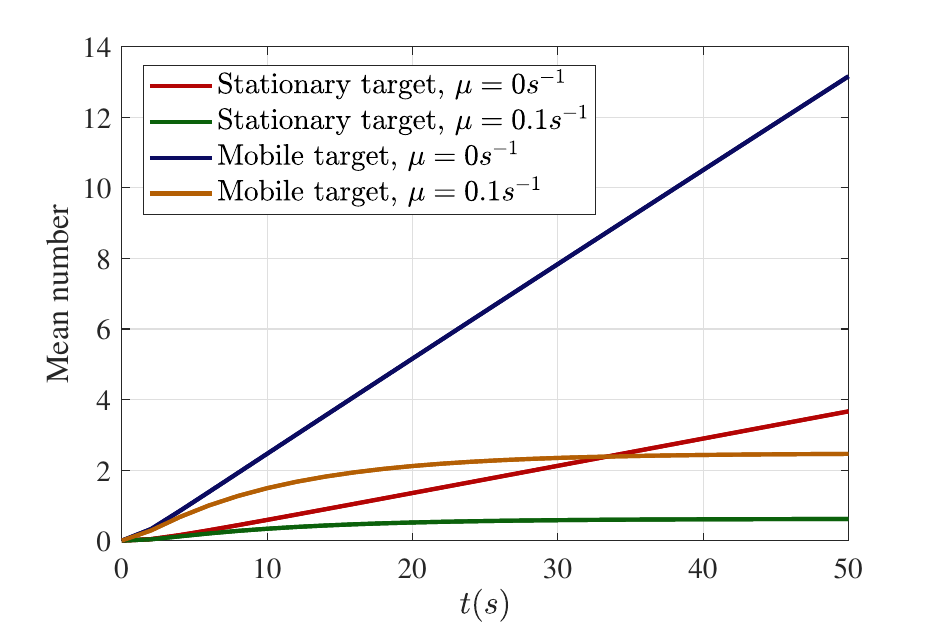}
	\caption{Analytical results showing the variation of the mean number of NMs detecting a stationary target and a mobile target within time $t$. Parameters :  $ a_1=3\mu m,\ a_2=4\mu m, D_t=100\mu m^2/s, D_1=100\mu m^2/s,\ D_2=75\mu m^2/s, \ \lambda_1=\lambda_2=1\times 10^{-5} \text{NMs}/\mu m^3,\ \text{ and \ } r=30\mu m$.}
	\label{fig:MeanNum}
\end{figure}
\textbf{\textit{Variation in Mean Number of NMs Detecting Target Over Time: }} 
The dynamics of the mean number of NMs detecting stationary versus moving targets within time $t$ is shown in Fig. \ref{fig:MeanNum}. For our analysis, we consider targets in two scenarios: without degradation (as per \eqref{nodegmean}) and with degradation (as in \eqref{degmean}). The mean number of NMs detecting the target increases when the target is mobile, as shown in Fig. \ref{fig:MeanNum}. Since more NMs pass through the area where the target is located as time passes, the average number of NMs that detect the target within time $t$ increases. In contrast, a target with degradation may eventually perish and cease to exist, making it impossible for NMs to detect it. This results in a constant mean number of NMs detecting the target within time $t$ when $t$ is significant, as shown in Fig. \ref{fig:MeanNum}.

\begin{figure}
	\centering
	\includegraphics[width=\linewidth]{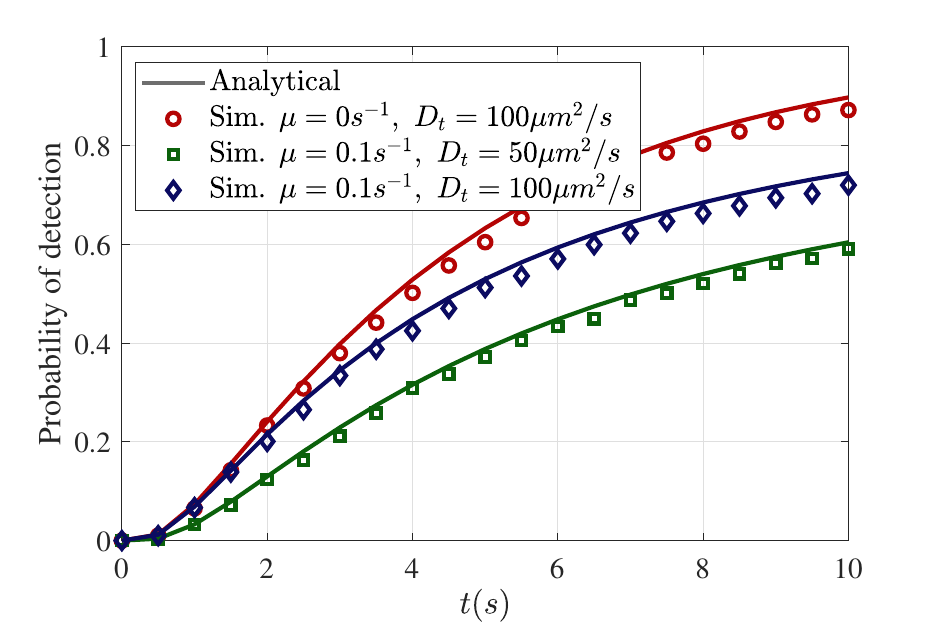}
	\caption{Probability of detecting a mobile target within time $t$ versus time. Parameters :  $ a_1=3\mu m,\ a_2=4\mu m, D_t=100\mu m^2/s, D_1=100\mu m^2/s,\ D_2=75\mu m^2/s, \ \lambda_1=\lambda_2=1\times 10^{-5} \text{NMs}/\mu m^3,\ \text{ and \ } r=30\mu m$.}
	\label{fig:MobTarget}
\end{figure}
\textbf{ \textit{Impact of Target Mobility on Detection Probability:}}
The variation of the detection probability of degradable/non-degradable and mobile target versus time for different values of diffusion coefficient $D_i$'s of NM is shown in Fig. \ref{fig:MobTarget}. Note that a higher diffusion coefficient implies a higher mobility of NMs. Since the target molecule
moves more quickly through the environment, the chances that it intersects an NM increases. Therefore, we can observe that the probability of detection increases as the target's diffusion coefficient increases. Such mobility will be beneficial when the target degradation rate is higher, as seen in Fig. \ref{fig:MobTarget}.

\textbf{\textit{Validation of Analytical Detection Probabilities through Particle-Based Simulations:}} 
Using particle base simulations, Fig. \ref{fig:cdf1} to Fig. \ref{fig:MobTarget} confirm the accuracy of the derived detection probability equations of target detection systems with stationary and mobile targets. We can confirm that analytical results match well with particle-based simulations from the aforementioned figures. Additionally, the utilization of analytical expressions significantly decreases the computational time required for particle-based simulations, reducing it from several hours to mere fractions of seconds.

\begin{figure}
	\centering
	\includegraphics[width=\linewidth]{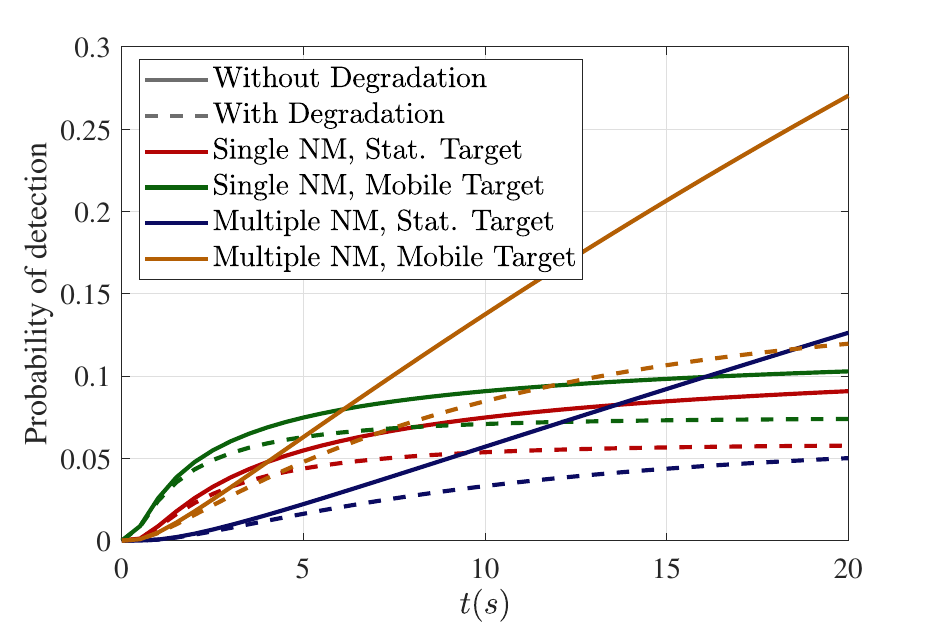}
	\caption{Comparison of probability of detection of the target for systems with stationary and mobile targets. Parameters for  single NMs case:  $  a_1=4\mu m,\ D_1=100\mu m^2/s ,\ d=30\mu m, \mu=0.1s^{-1}$. Parameters for multiple NM case:  $ a_1=3\mu m,\ a_2=4\mu m, D_t=100\mu m^2/s, D_1=100\mu m^2/s,\ D_2=75\mu m^2/s, \ \lambda_1=\lambda_2=1\times 10^{-6} \text{NMs}/\mu m^3,\ r=30\mu m,\ \text{ and \ } \mu=0.1s^{-1}$.}
	\label{fig:ProbAll}
\end{figure}
\textbf{ \textit{Comparative Analysis of Detection Probabilities of Different Systems:}}
 The comparison of various target detection strategies for non-degradable/degradable targets is shown in Fig. \ref{fig:ProbAll}. The target detection systems with multiple NMs have a higher detection probability than systems with a single NM for large enough $t$. Note that in a single NM case, the NM is placed at $d=r$, in multiple NM case, NMs are deployed outside $r$. The nearest NM can potentially be very far than $r$  in multiple NM case for low NM density. Due to this, in this scenario, the detection probability of a single NM can be better than that of a system with multiple NMs when $t$ is small. Via extensive simulations (not included here due to lack of space), we have observed that deploying multiple NMs is always better if the NM density is sufficiently high. We can also see from Fig. \ref{fig:ProbAll} that the probability of detection is higher when the target is mobile. However, when the target is degradable, the probability of detection decreases. 

\begin{figure}
	\centering
	\includegraphics[width=\linewidth]{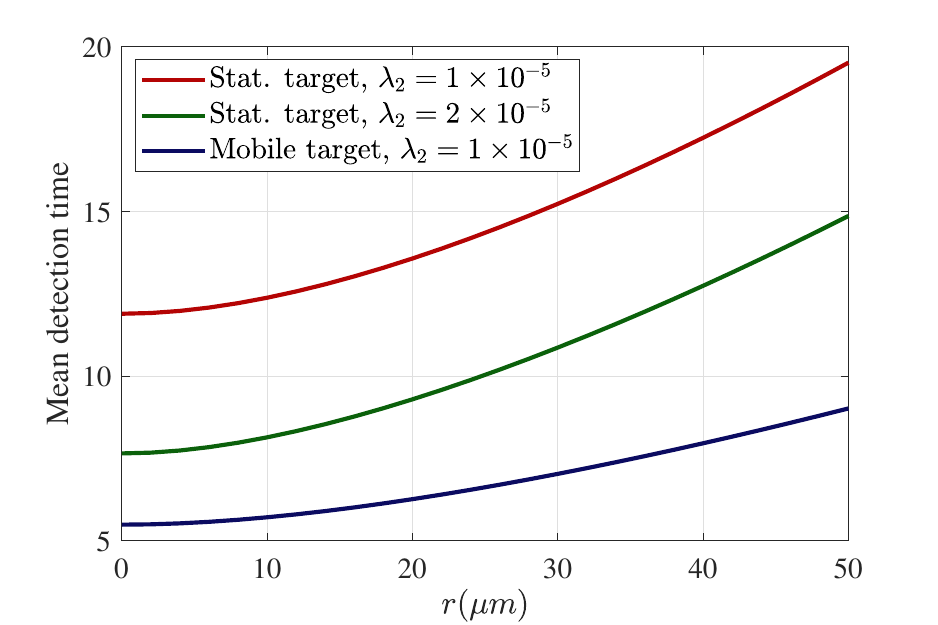}
	\caption{Comparison of mean detection time of the non-degrading target for systems with stationary and mobile targets. Parameters :  $ a_1=3\mu m,\ a_2=4\mu m, D_t=100\mu m^2/s, D_1=100\mu m^2/s,\ D_2=75\mu m^2/s,\ \text{ and \ } \lambda_1=1\times 10^{-5} \text{NMs}/\mu m^3$.}
	\label{fig:mdtAll}
\end{figure}
\textbf{ \textit{Mean Detection Time Analysis:}}
Fig. \ref{fig:mdtAll} illustrates the differences in mean detection time for systems with stationary and mobile targets  without degradation. As the value of $r$ increases, the initial spatial distribution of the NMs is further from the target, leading to a longer mean detection time due to the increased distance NMs need to travel to detect the target.
The mean detection time for a stationary target is longer than that for a mobile target. Additionally, as the density of NMs increases, so does the mean detection time. This is because a higher density means more NMs are present in the medium.
The parameter $\mu$ also plays a significant role. When $\mu$ is non-zero, the mean detection time can significantly increase. This is because there is a higher chance that the target will degrade before being detected by NMs, especially if we wait for an extended period. This will make detection time to be infinite.

\begin{figure}
	\centering
	\includegraphics[width=\linewidth]{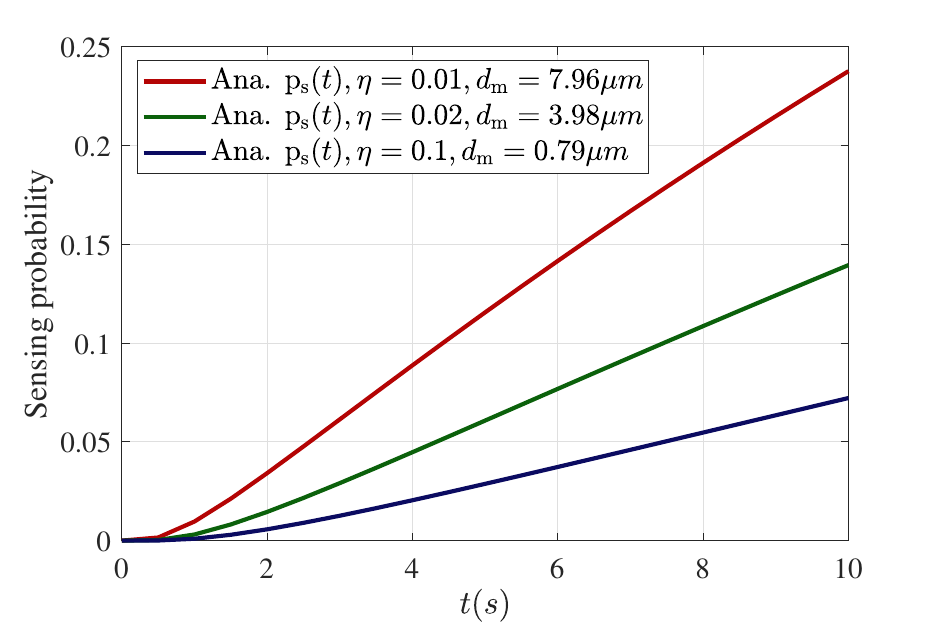}
	\caption{Probability of sensing versus time for different values of $\eta$. Parameters for multiple NMs case:  $  a_1=3\mu m,\ a_2=4\mu m,\ D_1=100\mu m^2/s,\ D_2=75\mu m^2/s, \ \lambda_1=\lambda_2=1\times 10^{-6} \text{NMs}/\mu m^3,\ \text{ and \ } r=30\mu m $.}
	\label{fig:detection}
\end{figure}
\textbf{\textit{Sensing Probability with Varying Thresholds:}}
Fig. \ref{fig:detection}  shows the change in sensing probability over time for various $ \eta $ values for a system with indirect sensing. The sensing probability decreases as the threshold $ \eta $ is increased. With the increase in $\eta$, the maximum distance up to which the concentration remains above a threshold $ \eta $ reduces, \ie $ d_m $ decreases. The area of sensing shrinks as $ d_m $ decreases, which also reduces the probability of sensing.

\begin{figure}
	\centering
	\includegraphics[width=\linewidth]{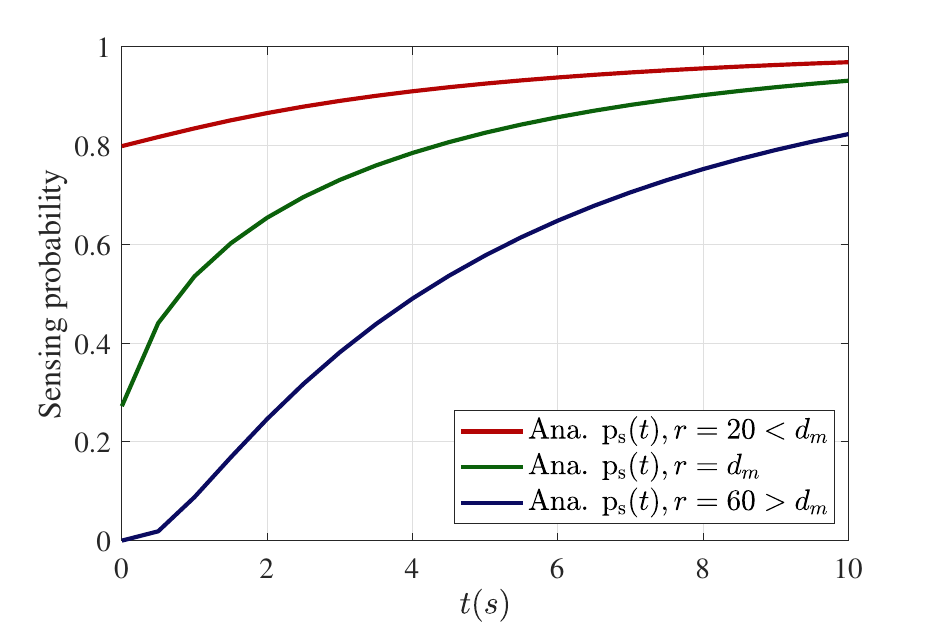}
	\caption{Probability of sensing versus time for different values of $r$. Parameters :  $  a_1=3\mu m,\ a_2=4\mu m,\ D_1=100\mu m^2/s,\ D_2=75\mu m^2/s, \ \lambda_1=\lambda_2=1\times 10^{-6} \text{NMs}/\mu m^3,\  \eta=0.002, \ \text{ and \ }d_{\mathrm{m}}=39.79\mu m $.}
	\label{fig:detection1}
\end{figure}
\textbf{\textit{Sensing Probability with Different Initial NM Distributions:}}
The variation of sensing probability with time for various values of $ r $ is shown in Fig. \ref{fig:detection1}. The NMs will initially be placed in the medium farther away from the target as $ r $ rises. In order to cross the  $ \ball{0}{d_\m} $ (region with concentration above detection threshold $ \eta $), NMs must therefore travel a greater distance. This will lower the probability of sensing. Note that, even with $r<d_{\mathrm{m}}$ the sensing probability is not $1$. This can be attributed to the sparse density of NMs, which might result in their nearest NM being deployed further than the specified 
$r$ distance.

\section{Conclusions}
\color{black}
In this work, we presented a comprehensive analytical framework for modeling and analyzing the performance of a target detection system utilizing a network of NMs of varying sizes. The investigation considered a broad spectrum of scenarios, including direct contact detection and indirect sensing, as well as the consideration of degradable and non-degradable target molecules, stationary and mobile targets.

Firstly, we presented the detection probability for a single NM  interacting with stationary and mobile target molecules, addressing both degradable and non-degradable targets. Then, we considered scenarios involving multiple mobile NMs deployed in the medium, accounting for stationary target molecules, with and without degradation. We derived the detection probabilities of such systems. The use of a PPP for NM deployment added realism to our model. Additionally, we explored the detection probability for systems with multiple mobile NMs and both degradable and non-degradable mobile targets and derived detection probabilities. We also derive the mean detection time, which refers to the average time it takes for the NMs, which are randomly deployed within a given environment, to identify a target molecule. Finally, we introduced an indirect sensing system where marker molecules emitted by the target are continuously monitored, and our analysis derived the sensing probability for such a system, offering an alternative approach to target detection.

Our analysis yields numerous insights, including the impact of target mobility, degradation, and NMs size on the probability of detection and sensing. Furthermore, we deduced that the adverse effects of target degradation can be mitigated by enhancing either the mobility or the size of the NMs. Another notable advantages of our approach is the substantial reduction in simulation time compared to particle-based simulations, rendering it a practical tool for efficiently studying these intricate systems. Overall, this research lays a foundation for further exploration of nanomachine-based target detection systems, holding promise for applications in healthcare, environmental monitoring, and security, among others.

\appendices

\section{Derivation of Lemma \ref{lmean}}\label{amean}
The mean number of NMs of radius $ a_i $ detecting the target at the origin is given by 
\begin{align}
\kappa_i(D_i,t)&=\expect{\bt{\x_i,t}}{N_{\Gamma_i}}\nonumber\\
&=\expect{\bt{\x_i,t}}{\sum_{\x_{ij}}\indc\left(\x_{ij} +\Sij(t)\cap\origin\neq \phi\right)}
\nonumber\\
&=\expect{\bt{\x_i,t}}{\sum_{\x_{ij}}\indc\left(\origin\in \x_i+\Sij(t) \right)}.
\end{align}
Now, applying Campbell theorem for marked point process \cite{andrews2023a},
\begin{align}
	\kappa_i(D_i,t)
	&=\lambda_i\expect{\bt{\x_i,t}}{\int_{\rthree/\ball{\origin}{r}}\indc\left(\origin\in \x_i+\Sij(t) \right)\diff \x_i}.
\end{align}
Noting that $ \Sij(t)= b_{ij}(t) \oplus \ball{0}{a_i} $, we get
\begin{align}
	\kappa_i(D_i,t)
		&=\lambda_i\mathbb{E}_{\bt{\x_i,t}}\left[\int_{\rthree/\ball{\origin}{r}}\indc\left(-\x_i\in\right.\right.\nonumber\\
	&\qquad\left.  b_{ij}(t)\oplus\ball{0}{a_i} \right)\diff \x_i\bigg]\nonumber\\
	&= \lambda_i\int_{\rthree/\ball{\origin}{r}} \mathbb{P}\left[  b_{ij}(t)\cap \ball{-\x_i}{a_i} \neq \phi\right] \diff \x_i.\label{aeq20}
\end{align}
Note that, $ \mathbb{P}\left[  b_{ij}(t)\cap \ball{-\x_i}{a_i} \neq \phi\right] $ is the probability that a molecule emitted from origin hits a sphere of radius $ a_i $ with center located at $ -\x_i $ within time $ t $. From \cite{yilmaz2014a}, 
\begin{align}
	\mathbb{P}\left[  b_{ij}(t)\cap \ball{-\x_i}{a_i} \neq \phi\right]=\frac{a_i}{\|\x_i\|}\erfc\left(\frac{\|\x_i\|-a_i}{\sqrt{4D_it}}\right).\label{3rxold}
\end{align}
Now, substituting \eqref{3rxold} in \eqref{aeq20} gives
\begin{align}
	\kappa_i(D_i,t)= 4\pi a_i\lambda_i\int_{r}^{\infty} r_i\erfc\left(\frac{r_i-a_i}{\sqrt{4D_it}}\right) \diff r_i.\label{aeq21}
\end{align}
Solving \eqref{aeq21} further gives Lemma \ref{lmean}.
%%%%%%%%%%%%%%%%%%%%%%%%%%%%%%%%%%%%%%%%%
\section{Derivation of Theorem \ref{th1}}\label{ath1}
The probability that the NMs in $\Phi_i$ do not detect the target at the origin is the same as the probability that $N_{\Gamma_i}= 0$. This probability is given by
\begin{align}
	\Probi{t}&=\mathbb{P}\left[N_{\Gamma_i}= 0\right]\nonumber\\
	&=\exp(-\kappa_i(D_i,t)).
\end{align}
Here, the above equation is due to the void probability of PPP $\Gamma_i$ \cite{andrews2023a}. 

Now the probability that any of the NMs in $ \Phi $ detect the target molecule within time $ t $ is given by
\begin{align}
	\Prob{t}&=1-\prod_{i=1}^{n}\Probi{t}=1-\exp\left(-\sum_{i=1}^{n}\kappa_i(D_i,t) \right).
\end{align}
%%%%%%%%
%The probability that there are $ n $ PPP distributed points of density $ \lambda $ inside a closed volume $ |\mathsf{C}| $ is given by
%\begin{align}
%	\Prob{n}=\exp\left(-\lambda|\mathsf{C}|\right)\frac{\left(-\lambda|\mathsf{C}|\right)^n}{n!}.
%\end{align}
%Note that, $ \lambda|\mathsf{C}| $ is the mean number of PPP distributed points inside $ \mathsf{C} $. Therefore, the probability that any of the NMs in $ \Phi_i $ do not detect the target molecule at the origin within time $ t $ is 
%\begin{align}
%	\Probi{t}&=\exp\left(-\expect{\Phi_i}{\mid \Sij(t)\mid}\right)\nonumber\\
%	&=\exp\left(2\pi a_i\lambda_i\left(a_i^2-r^2+2D_it\right)\erfc\left(\frac{r-a_i}{\sqrt{4D_it}}\right)\right.\nonumber\\
%	&\left.+4\lambda_i\sqrt{\pi D_it}a_i(r+a_i)\exp\left(-\frac{(r-a_i)^2}{4D_it}\right)\right).
%\end{align}
%Now the probability that any of the NMs in $ \Phi $ detect the target molecule within time $ t $ is given by
%\begin{align}
%	\Prob{t}&=1-\prod_{i=1}^{n}\Probi{t}\nonumber\\
%	&=1-\exp\left(-2\pi\sum_{i=1}^{n} a_i\lambda_i\left(a_i^2-r^2+2D_it\right)\erfc\left(\frac{r-a_i}{\sqrt{4D_it}}\right)\right.\nonumber\\
%	&\left.-4\sum_{i=1}^{n}\lambda_i\sqrt{\pi D_it}a_i(r+a_i)\exp\left(-\frac{(r-a_i)^2}{4D_it}\right)\right).
%\end{align}

\section{Derivation of Lemma \ref{ldmean}}\label{admean}
The mean number of NMs of radius $ a_i $ detecting the target at the origin before its degradation is given by
\begin{align}
	\expect{t_\diff}{\rho(D_i,t\mid t_\diff )}&=\kappa_i(D_i,t)\expect{t_\diff}{\indc\left(t_{\diff}>t\right)}\nonumber\\
	&\quad+\expect{t_\diff}{\kappa_i(t_\diff)\indc\left(t_{\diff}<t\right)}.
	%&=\expect{}{\underbrace{\int_{\rthree/\ball{\origin}{r}}\indc\left(-\x_i\in b_{ij}(\min\{t,t_\diff\})\oplus\ball{0}{a_i} \right)\diff \x_i}_{\rho(D_i,t\mid t_\diff )}}\nonumber\\
	%	&=\lambda_i\expect{t_\diff}{\int_{\rthree/\ball{\origin}{r}}\indc\left(-\x_i\in b_{ij}(t)\oplus\ball{0}{a_i} \right)\indc\left(t_{\diff}>t\right)\diff \x_i}\nonumber\\
	%	& \quad+\lambda_i\expect{t_\diff}{\int_{\rthree/\ball{\origin}{r}}\indc\left(-\x_i\in b_{ij}(t)\oplus\ball{0}{a_i} \right)\indc\left(t_{\diff}<t\right)\diff \x_i}
	\label{eqfirst}
\end{align} 
%\begin{align}
%	&\expect{t_\diff}{\rho(D_i,t\mid t_\diff )}\nonumber\\	
%	%&=\expect{}{\underbrace{\int_{\rthree/\ball{\origin}{r}}\indc\left(-\x_i\in b_{ij}(\min\{t,t_\diff\})\oplus\ball{0}{a_i} \right)\diff \x_i}_{\rho(D_i,t\mid t_\diff )}}\nonumber\\
%&=\lambda_i\expect{t_\diff}{\int_{\rthree/\ball{\origin}{r}}\indc\left(-\x_i\in b_{ij}(t)\oplus\ball{0}{a_i} \right)\indc\left(t_{\diff}>t\right)\diff \x_i}\nonumber\\
%	& \quad+\lambda_i\expect{t_\diff}{\int_{\rthree/\ball{\origin}{r}}\indc\left(-\x_i\in b_{ij}(t)\oplus\ball{0}{a_i} \right)\indc\left(t_{\diff}<t\right)\diff \x_i}\label{eqfirst}
%\end{align}
The first term of \eqref{eqfirst} gives $ \kappa_i(D_i,t)\exp(\mu t) $, because $\expect{t_\diff}{\indc\left(t_{\diff}>t\right)}=\exp(\mu t)$.

Substituting the value from \eqref{nodegmean}, the second term becomes
{\begin{align}
	&\expect{t_\diff}{\kappa_i(t_\diff)\indc\left(t_{\diff}<t\right)}=
	e^{-(r-a_i)\sqrt{\frac{\mu}{D_i}}}\erfc\left(\frac{r-a_i}{\sqrt{4D_it}}-\sqrt{\mu t}\right)\nonumber\\
	&\qquad\times 2\pi\lambda_ia_i\left[\frac{D_i}{\mu}+r\sqrt{\frac{D_i}{\mu}}\right] +e^{(r-a_i)\sqrt{\frac{\mu}{D_i}}} \nonumber\\
	&\qquad\times\erfc\left(\frac{r-a_i}{\sqrt{4D_it}}+\sqrt{\mu t}\right) 2\pi\lambda_ia_i\left[\frac{D_i}{\mu}-r\sqrt{\frac{D_i}{\mu}}\right]\nonumber\\
	&\qquad-4\pi a_i\lambda_i\frac{D_i}{\mu}e^{-\mu t}\erfc\left(\frac{r-a_i}{\sqrt{4D_it}}\right)- \kappa_i(D_i,t)\exp(\mu t).
\end{align}}
Substituting the first and second terms in \eqref{eqfirst}, and solving further gives \eqref{degmean}.
\section{Derivation of Theorem \ref{thdeg}}\label{athdeg}
Similar to Appendix \ref{ath1}, condition on $t_\diff$, the probability that any of the NMs in $ \Phi_i $ do not detect the degradable target molecule at the origin within time $ t $ can be derived as
\begin{align}
	\Probi{t\mid t_\diff}	&=\exp\left(-\lambda_i\times\right.\nonumber\\
	&\qquad\left.\expect{\Sij(t)\mid t_\diff}{|\x_{ij}+\Sij(\min\{t,t_\diff\})|}\right)\nonumber\\
		&=\exp\left(-\rho(D_i,t\mid t_\diff )\right).\label{pieq}
\end{align}

Conditioned on $t_\diff$, since the event of detection of the degradable target by NMs in each of the $\Phi_i$ are independent, the probability that any of the NMs detect the target molecule before its degradation within time $ t $ is given by
\begin{align}
	\Probd{t\mid t_\diff}&=1-\prod_{i=1}^{n}\Probi{t\mid t_\diff}.
\end{align}
Hence, averaging over $t_\diff$, we get
	\begin{align}
		\Probd{t}&=1-\expect{t_{\diff}}{\exp\left(-\sum_{i=1}^{n}\rho(D_i,t\mid t_\diff )\right)}.
	\end{align}
\section{Derivation of Corollary \ref{coronew}}\label{athdegcoro}
Using cumulant expansion \cite{hald2000a} in \eqref{th4prob} gives
\begin{align}
	\Probd{t}&=1-\exp\left(\sum_{k=1}^{\infty}\frac{\alpha_k(-1)^k}{k!}\right),\label{cumu}
\end{align}
where $ \alpha_k $ is the $ k - $ th cumulant of $\sum_{i=1}^{n}\rho(D_i,t\mid t_\diff )$.

 Note that, the first cumulant is the mean with respect to $ t_{\diff} $ (which is $ \sum_{i=1}^{n}\expect{}{\rho(D_i,t\mid t_\diff )} $), second cumulant is the variance and third cumulant is the third central mean of the random variable $ \sum_{i=1}^{n}\rho(D_i,t\mid t_\diff ) $.\par
Now, approximating \eqref{cumu} by retaining the first cumulant and removing the higher order cumulants gives
\begin{align}
	\Probd{t}&\approx1-\exp\left(-\alpha_1\right)=1-\exp\left(-\sum_{i=1}^{n}\expect{}{\rho(D_i,t\mid t_\diff )}\right),
\end{align}
which is \eqref{approxeq}. 
\section{Derivation of Theorem \ref{th1}}\label{amob}
Let us denote the event of detection of the target by any of the $ \NMij $ in $ \Phi_i $ by $ \Evei $. \ie
\begin{align*}
	\Evei&=\cup_{\x_{ij}\in\Phi_i}\left\{\origin\in \x_{ij}+\Tij(t)\right\}\nonumber\\
	&=\left(\cap_{\x_{ij}\in\Phi_i}\left\{\origin\notin \x_{ij}+\Tij(t)\right\}\right)^c.
\end{align*}
Therefore, the probability that any of the NMs in $ \Phi_i $ do not detect the target molecule at the origin within time $ t $ is 
\begin{align}
	\Probi{t}&=1-\mathbb{P}[\Evei]=\expect{}{\prod_{\x_{ij}\in\Phi_i}\indc\left(\origin\notin \x_{ij}+\Tij(t)\right)}\nonumber\\
	&\stackrel{(a)}{=}\mathbb{E}_{\bt{\x_i,t}}\left[\exp\left(-\lambda_i\int_{\rthree/\ball{\origin}{r}}\left(1-\indc\left(\origin\notin\right.\right.\right.\right.\nonumber\\
	&\qquad \left.\left.\left. \x_{i}+\Tij(t)\right)\right)\diff \x_i\bigg)\right]\nonumber\\
		&=\mathbb{E}_{\bt{\x_i,t}}\left[\exp\left(-\lambda_i\int_{\rthree/\ball{\origin}{r}}\indc\left(\origin\in\right.\right.\right.\nonumber\\
		&\qquad \left.\left. \x_i+\Tij(t) \right)\diff \x_i\right)\bigg]\nonumber\\
		&=\mathbb{E}_{\bt{\x_i,t}}\left[\exp\left(-\lambda_i\mid \x_i+\Tij(t)\mid \right)\diff \x_i\right)\bigg]\nonumber\\
\end{align}
Here, $ (a) $ is due to the probability generating functional (PGFL) of the PPP \cite{andrews2023a}.
Using cumulant expansion as shown in \eqref{cumu} in the above equation and retaining the first cumulant gives,
\begin{align}
	\Probi{t}&\approx\exp\left(-\lambda_i\mathbb{E}_{\bt{\x_i,t}}\left[\mid \x_i+\Tij(t)\mid\bigg] \right)\diff \x_i\right)\nonumber\\
	&\approx\exp(-\kappa(\diffeff{i},t))
\end{align}
 Now the approximate probability that any of the NMs in $ \Phi $ detect the target molecule within time $ t $ is given by
\begin{align}
	\Prob{t}&\approx1-\prod_{i=1}^{n}\Probi{t}=1-\exp\left(-\sum_{i=1}^{n}\kappa_i(\diffeff{i},t) \right).
\end{align}
The derivation of the detection probability of a mobile degradable target is similar to the derivation in Appendix \ref{athdegcoro}.
\section{Derivation of Theorem \ref{tsense1}}\label{asense1}
The mean number of NMs of radius $ a_i $ sensing the presence of the target at the origin at  time instant $ t $ is given by 
\begin{align}
	\Gamma(\Phi_i,t)= \expect{}{\sum_{\y_{ij}\in \Phi_i}\indc\left(\ball{\origin}{d_{\m}}\cap  \ball{\y_{ij}}{a_i}\neq \phi\right) },\label{ce1}
\end{align}
Applying Campbell's theorem on \eqref{ce1} gives
\begin{align}
	\Gamma(\Phi_i,t)&= \int \lambda(t,\y_i) \indc\left(\ball{\origin}{d_{\m}}\cap  \ball{\y_{i}}{a_i}\neq \phi\right)\diff \y_i\nonumber\\
	&= \int \lambda(t,\y_i) \indc\left(\y_{i}\in \ball{\origin}{a_i+d_{\m}} \right)\diff \y_i\nonumber\\
	&=\int_{\ball{\origin}{a_i+d_{\m}}} \lambda(t,\y_i) \diff \y_i.
\end{align}
The probability that the NMs in $\Phi_i$ do not sense the target is the same as the probability that $\Gamma(\Phi_i,t)= 0$. This probability is given by
\begin{align}
	\Probi{t}&=\mathbb{P}\left[\Gamma(\Phi_i,t)= 0\right]\nonumber\\
	&=\exp(-\Gamma(\Phi_i,t)).
\end{align}
The probability of sensing the presence of the target by any of the NMs at any time instant is given by
\begin{align}
	\Prob{t}&=1-\prod_{i=1}^{n}\Probi{t}\nonumber\\
	&=1-\prod_{i=1}^{n}\exp\left(-\int_{\ball{\origin}{a_i+d_{\m}}} \lambda(t,\y_i) \diff \y_i\right).
\end{align}
\bibliographystyle{IEEEtran}
\bibliography{nanorefn1}

% Generated by IEEEtran.bst, version: 1.14 (2015/08/26)
\begin{thebibliography}{10}
\providecommand{\url}[1]{#1}
\csname url@samestyle\endcsname
\providecommand{\newblock}{\relax}
\providecommand{\bibinfo}[2]{#2}
\providecommand{\BIBentrySTDinterwordspacing}{\spaceskip=0pt\relax}
\providecommand{\BIBentryALTinterwordstretchfactor}{4}
\providecommand{\BIBentryALTinterwordspacing}{\spaceskip=\fontdimen2\font plus
\BIBentryALTinterwordstretchfactor\fontdimen3\font minus
  \fontdimen4\font\relax}
\providecommand{\BIBforeignlanguage}[2]{{%
\expandafter\ifx\csname l@#1\endcsname\relax
\typeout{** WARNING: IEEEtran.bst: No hyphenation pattern has been}%
\typeout{** loaded for the language `#1'. Using the pattern for}%
\typeout{** the default language instead.}%
\else
\language=\csname l@#1\endcsname
\fi
#2}}
\providecommand{\BIBdecl}{\relax}
\BIBdecl

\bibitem{nakano2013a}
T.~Nakano, A.~W. Eckford, and T.~Haraguchi, \emph{Molecular {{Communication}}},
  1st~ed.\hskip 1em plus 0.5em minus 0.4em\relax {Cambridge University Press},
  Sep. 2013.

\bibitem{nakano2012}
T.~Nakano, M.~J. Moore, {Fang Wei}, A.~V. Vasilakos, and {Jianwei Shuai},
  ``Molecular communication and networking: {{Opportunities}} and challenges,''
  \emph{IEEE Trans. Nanobioscience}, vol.~11, no.~2, pp. 135--148, Jun. 2012.

\bibitem{kuran2010}
M.~{\c S}. Kuran, H.~B. Yilmaz, T.~Tugcu, and B.~{\"O}zerman, ``Energy model
  for communication via diffusion in nanonetworks,'' \emph{Nano Communication
  Networks}, vol.~1, no.~2, pp. 86--95, Jun. 2010.

\bibitem{okaie2016a}
Y.~Okaie, T.~Nakano, T.~Hara, and S.~Nishio, \emph{Target {{Detection}} and
  {{Tracking}} by {{Bionanosensor Networks}}}, ser. {{SpringerBriefs}} in
  {{Computer Science}}.\hskip 1em plus 0.5em minus 0.4em\relax {Singapore}:
  {Springer Singapore}, 2016.

\bibitem{nakano2017a}
T.~Nakano, Y.~Okaie, S.~Kobayashi, T.~Koujin, C.-H. Chan, Y.-H. Hsu, T.~Obuchi,
  T.~Hara, Y.~Hiraoka, and T.~Haraguchi, ``Performance evaluation of
  leader\textendash follower-based mobile molecular communication networks for
  target detection applications,'' \emph{IEEE Trans. Commun.}, vol.~65, no.~2,
  pp. 663--676, Feb. 2017.

\bibitem{mosayebi2018}
R.~Mosayebi, W.~Wicke, V.~Jamali, A.~Ahmadzadeh, R.~Schober, and
  M.~{Nasiri-Kenari}, ``Advanced target detection via molecular
  communication,'' in \emph{Proc. {{GLOBECOM}}}, Dec. 2018, pp. 1--7.

\bibitem{sabu2020b}
N.~V. Sabu and A.~K. Gupta, ``Detection probability in a molecular
  communication via diffusion system with multiple fully-absorbing receivers,''
  \emph{IEEE Commun. Lett.}, vol.~24, no.~12, pp. 2824--2828, Dec. 2020.

\bibitem{andrews2023a}
J.~G. Andrews, A.~K. Gupta, A.~Alammouri, and H.~S. Dhillon, \emph{An
  Introduction to Cellular Network Analysis Using Stochastic Geometry}, ser.
  Synthesis {{Lectures}} on {{Learning}}, {{Networks}}, and
  {{Algorithms}}.\hskip 1em plus 0.5em minus 0.4em\relax {Cham}: {Springer
  International Publishing}, 2023.

\bibitem{haenggi2012}
M.~Haenggi, \emph{Stochastic Geometry for Wireless Networks}, 1st~ed.\hskip 1em
  plus 0.5em minus 0.4em\relax {Cambridge University Press}, Oct. 2012.

\bibitem{sabu2019}
N.~V. Sabu and A.~K. Gupta, ``Analysis of diffusion based molecular
  communication with multiple transmitters having individual random information
  bits,'' \emph{IEEE Trans. Mol. Biol. Multi-Scale Commun.}, vol.~5, no.~3, pp.
  176--188, Dec. 2019.

\bibitem{deng2017a}
Y.~Deng, A.~Noel, W.~Guo, A.~Nallanathan, and M.~Elkashlan, ``Analyzing
  large-scale multiuser molecular communication via 3-{{D}} stochastic
  geometry,'' \emph{IEEE Trans. Mol. Biol. Multi-Scale Commun.}, vol.~3, no.~2,
  pp. 118--133, Jun. 2017.

\bibitem{dissanayake2019a}
M.~B. Dissanayake, Y.~Deng, A.~Nallanathan, M.~Elkashlan, and U.~Mitra,
  ``Interference mitigation in large-scale multiuser molecular communication,''
  \emph{IEEE Trans. Commun.}, vol.~67, no.~6, pp. 4088--4103, Jun. 2019.

\bibitem{cerny2008}
R.~{\v C}ern{\'y}, S.~Funken, and E.~Spodarev, ``On the boolean model of wiener
  sausages,'' \emph{Methodol Comput Appl Probab}, vol.~10, no.~1, pp. 23--37,
  Mar. 2008.

\bibitem{heren2015a}
A.~C. Heren, H.~B. Yilmaz, C.-B. Chae, and T.~Tugcu, ``Effect of degradation in
  molecular communication: {{Impairment}} or enhancement?'' \emph{IEEE Trans.
  Mol. Biol. Multi-Scale Commun.}, vol.~1, no.~2, pp. 217--229, Jun. 2015.

\bibitem{yilmaz2014a}
H.~B. Yilmaz, A.~C. Heren, T.~Tugcu, and C.-B. Chae, ``Three-dimensional
  channel characteristics for molecular communications with an absorbing
  receiver,'' \emph{IEEE Commun. Lett.}, vol.~18, no.~6, pp. 929--932, Jun.
  2014.

\bibitem{hald2000a}
A.~Hald, ``The early history of the cumulants and the {{Gram-Charlier}}
  series,'' \emph{Int Statistical Rev}, vol.~68, no.~2, pp. 137--153, Aug.
  2000.

\bibitem{huang2018}
S.~Huang, L.~Lin, H.~Yan, J.~Xu, and F.~Liu, ``Mean and variance of received
  signal in diffusion-based mobile molecular communication,'' in \emph{Proc.
  {{GLOBECOM}}}, Dec. 2018, pp. 1--6.

\bibitem{nakano2019}
T.~Nakano, Y.~Okaie, S.~Kobayashi, T.~Hara, Y.~Hiraoka, and T.~Haraguchi,
  ``Methods and applications of mobile molecular communication,'' \emph{Proc.
  IEEE}, vol. 107, no.~7, pp. 1442--1456, Jul. 2019.

\bibitem{jamali2019}
V.~Jamali, A.~Ahmadzadeh, W.~Wicke, A.~Noel, and R.~Schober, ``Channel modeling
  for diffusive molecular communication\textemdash{{A}} tutorial review,''
  \emph{Proc. IEEE}, vol. 107, no.~7, pp. 1256--1301, Jul. 2019.

\bibitem{bossert1963a}
W.~H. Bossert and E.~O. Wilson, ``The analysis of olfactory communication among
  animals,'' \emph{Journal of Theoretical Biology}, vol.~5, no.~3, pp.
  443--469, Nov. 1963.

\end{thebibliography}
\end{document}